\documentclass[fleqn,useAMS,usenatbib]{mnras}
\usepackage{mathptmx}
\usepackage{color}
\usepackage{graphics,graphicx,amssymb,amsmath,natbib}
\usepackage{float}
\usepackage{color}
\usepackage{lscape,graphicx}
\usepackage{amsmath}
\usepackage{amssymb}
\usepackage{multirow}
\usepackage{afterpage}
\usepackage{threeparttable}
\usepackage{subfigure}
\usepackage{rotating}
\usepackage{lscape}
\usepackage{caption}
\usepackage{verbatim}
\usepackage{morefloats}
\usepackage[T1]{fontenc}
\usepackage{ae,aecompl}
\usepackage{graphicx}
\usepackage{epstopdf}
\usepackage{amsmath}
\usepackage{amssymb}
\usepackage{enumerate}
\usepackage{dblfloatfix}
\usepackage{lipsum}
\usepackage[compact]{titlesec}
\titleformat{\section}
  {\normalfont\fontsize{12}{15}\bfseries}{\thesection}{1em}{}

\def\cha    {{\em Chandra}\/}

\def\xmm        {{\em XMM-Newton}\/}
\def\XMM        {{\em XMM}\/}

\def\subaru        {{\em Subaru}\/}
\def\muse        {{\em MUSE}\/}

\def\ha        {H$\alpha$\/}
\newcommand{\km}{km\,s$^{-1}$}
\newcommand{\hi}{H\,{\sc i}}
\newcommand{\hii}{H\,{\sc ii}}

\usepackage{orcidlink} 
\newcommand{\orcidcg}{0000-0003-0628-5118} 
\newcommand{\orcidms}{0000-0001-5880-0703} 
\newcommand{\orcidlj}{0000-0001-7500-0660} 
\newcommand{\orcidrl}{0000-0003-4509-7822} 
\newcommand{\orcidwf}{0000-0002-9478-1682} 
\newcommand{\orcidcj}{0000-0003-2206-4243} 
\newcommand{\orcidmy}{0000-0001-7550-2281} 
\newcommand{\orcidmf}{0000-0002-9043-8764} 
\newcommand{\orcidpj}{0000-0002-1640-5657} 
\newcommand{\orcidec}{0000-0002-0322-884X} 
\newcommand{\orcidiz}{0000-0001-7630-8085} 
\newcommand{\orcidab}{0000-0002-9795-6433} 

\title[The BIG X-ray tail]{\vspace{-0.5cm}The BIG X-ray tail}
\author
[Ge et al.]{Chong Ge$^{1}$\thanks{chong.ge@uah.edu}\orcidlink{\orcidcg}, Ming Sun$^{1}$\thanks{ming.sun@uah.edu}\orcidlink{\orcidms}, Masafumi Yagi$^{2}$\orcidlink{\orcidmy}, Matteo Fossati$^{3,4}$\orcidlink{\orcidmf},
\newauthor
William Forman$^{5}$\orcidlink{\orcidwf},
Pavel J\'{a}chym$^{6}$\orcidlink{\orcidpj},
Eugene Churazov$^{7,8}$\orcidlink{\orcidec},
Irina Zhuravleva$^{9}$\orcidlink{\orcidiz}
\newauthor
Alessandro Boselli$^{10}$\orcidlink{\orcidab},
Christine Jones$^{5}$\orcidlink{\orcidcj},
Li Ji$^{11}$\orcidlink{\orcidlj}, and Rongxin Luo$^{1}$\orcidlink{\orcidrl}\\
$^{1}$Department of Physics and Astronomy, University of Alabama in Huntsville, Huntsville, AL 35899, USA\\
$^{2}$National Astronomical Observatory of Japan, 2-21-1, Osawa, Mitaka, Tokyo, 181-8588, Japan\\
$^{3}$Dipartimento di Fisica G. Occhialini, Universit\`{a} degli Studi di Milano Bicocca, Piazza della Scienza 3, I-20126 Milano, Italy \\
$^{4}$INAF-Osservatorio Astronomico di Brera, via Brera 28, I-20121 Milano, Italy \\
$^{5}$Harvard-Smithsonian Center for Astrophysics, 60 Garden Street, Cambridge, MA 02138, USA\\
$^{6}$Astronomical Institute of the Czech Academy of Sciences, Bo\v{c}n\'{i} II 1401, 141 00, Prague, Czech Republic\\
$^{7}$Max Planck Institute for Astrophysics, Karl-Schwarzschild-Str. 1, 85741, Garching, Germany\\
$^{8}$Space Research Institute (IKI), Profsoyuznaya 84/32, Moscow, 117997, Russia\\
$^{9}$Department of Astronomy and Astrophysics, The University of Chicago, Chicago, IL 60637, USA\\
$^{10}$Aix Marseille Universit\'{e}, CNRS, LAM (Laboratoire d’ Astro physique de Marseille) UMR 7326, 13388, Marseille, France \\
$^{11}$Purple Mountain Observatory, Chinese Academy of Sciences, Nanjing 210008, China
}
\begin{document}
\date{\vspace{-0.35cm}Accepted 2021 September 13. Received 2021 September 13; in original form 2021 August 27}

\pubyear{2021}

\maketitle
\vspace{-6cm}
\begin{abstract}
Galaxy clusters grow primarily through the continuous accretion of group-scale haloes. Group galaxies experience preprocessing
during their journey into clusters.
A star-bursting compact group, the Blue Infalling Group (BIG), is plunging into the nearby cluster A1367. Previous optical observations reveal rich tidal features in the BIG members, and a long \ha\ trail behind.
Here we report the discovery of a projected $\sim 250$ kpc X-ray tail behind the BIG using \cha\ and \xmm\ observations. 
The total hot gas mass in the tail is $\sim 7\times 10^{10}\ {\rm M}_\odot$ with an X-ray bolometric luminosity of $\sim 3.8\times 10^{41}$ erg s$^{-1}$. The temperature along the tail is $\sim 1$ keV, but the apparent metallicity is very low, an indication of the multi-$T$ nature of the gas. The X-ray and \ha\ surface brightnesses in the front part of the BIG tail follow the tight correlation established from a sample of stripped tails in nearby clusters, which suggests the multiphase gas originates from the mixing of the stripped interstellar medium (ISM) with the hot intracluster medium (ICM).
Because thermal conduction and hydrodynamic instabilities are significantly suppressed, the stripped ISM can be long lived and produce ICM clumps.
The BIG provides us a rare laboratory to study galaxy transformation and preprocessing.
\end{abstract}

\begin{keywords}
galaxies: clusters: individual: Abell 1367 -- galaxies: clusters: intracluster medium -- galaxies: groups: general -- X-rays: galaxies: clusters 
\end{keywords}

\section{INTRODUCTION} \label{sec:intro}
Galaxies evolve during hierarchical structure formation.
The key to galaxy evolution is gas removal, which can be caused by internal and external mechanisms.
Internal mechanisms include star formation (SF) and active galactic nucleus (AGN) feedback.
External mechanisms include gravitational (e.g. galaxy-galaxy tidal interactions) and hydrodynamic interactions (e.g. ram pressure stripping; RPS).
Hybrid processes of gravitational and hydrodynamic interactions acting on galaxies referred to as `preprocessing', happen when galaxy groups fall into clusters (e.g. \citealt{2006PASP..118..517B}).

An actively star-bursting compact group
is infalling into the core of A1367, a dynamically young merging cluster (e.g. \citealt{1983ApJ...265...26B,2002ApJ...576..708S,2004A&A...425..429C}).
The group was discovered in the \ha\ survey of nearby clusters by \cite{2002ApJ...578..842S}. Its presence was also revealed in the \ha\ data by \cite{2002A&A...384..383I} and \cite{2003ApJ...597..210G} who named it the Blue Infalling Group (BIG).
\cite{2006A&A...453..847C} presented deep \ha\ imaging and multislit spectroscopy of the BIG, and suggested that the BIG experiences preprocessing.
They also found that \ha\ stream extends $>150$ kpc to the NW.
\cite{2017ApJ...839...65Y} reported that the \ha\ extends much further; the BIG+tail are $\sim 330$ kpc ($\sim 180$ kpc for tail only) in projection.
\cite{2019MNRAS.484.2212F} suggested the BIG is shaped by the mutual influence of gravitational interactions and RPS, based on the {\em VLT}/\muse\ observations.

\begin{figure}
\begin{center}
\centering
\includegraphics[angle=0,width=0.475\textwidth]{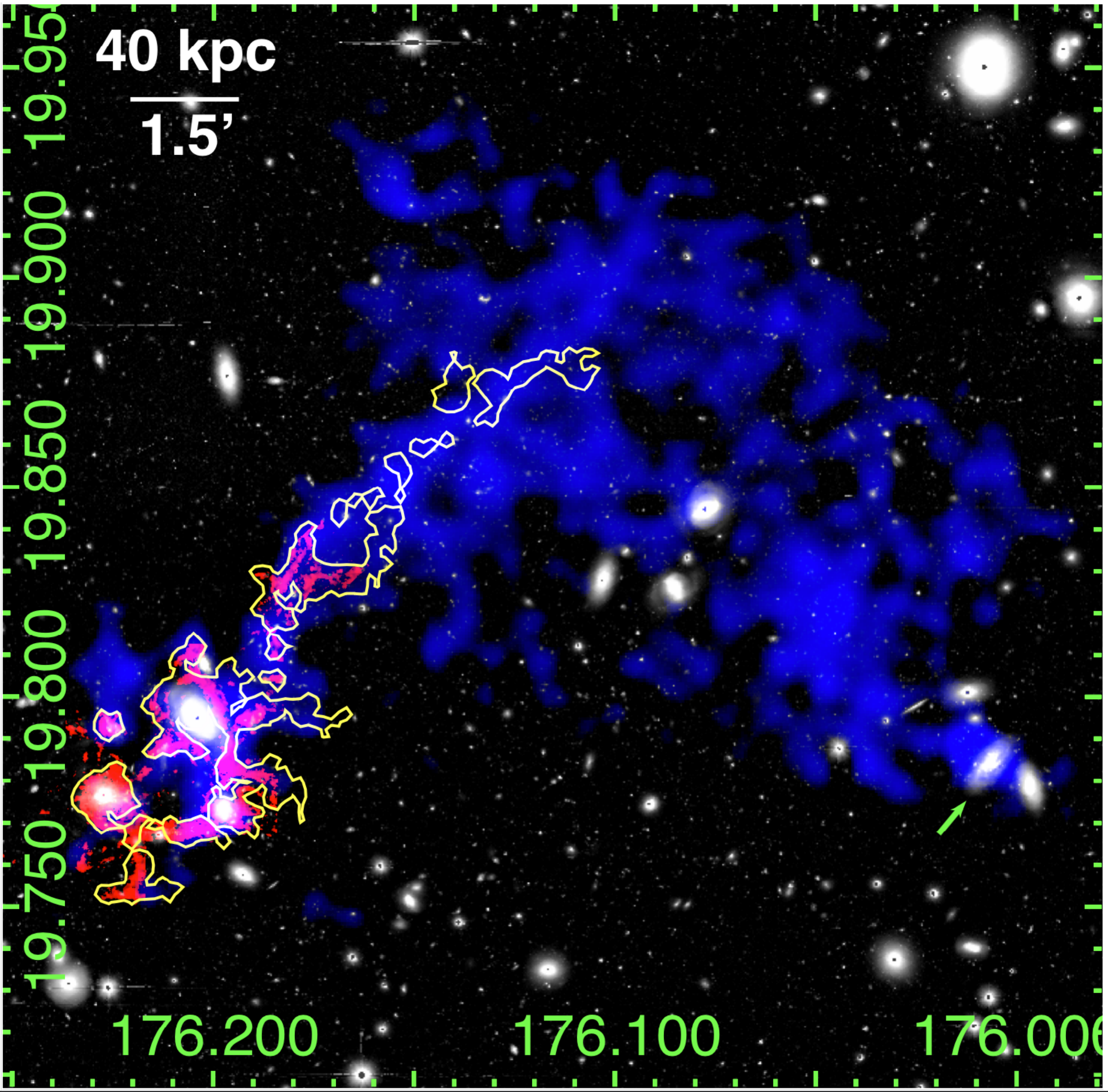}
\vspace{-0.5cm}
\caption{
Three-color composite X-ray/optical image around the Blue Infalling Group (BIG). White: \subaru\ $B$-band image; red: \muse\ H$\alpha$ image \citep{2019MNRAS.484.2212F}, the yellow contour is from the \subaru\ \ha\ image \citep{2017ApJ...839...65Y} with a larger coverage than \muse; blue: merged \cha\ and \XMM\ residual diffuse image. 
The green arrow points to another RPS galaxy CGCG~097-093.
The RA and Dec are in J2000.
}
\label{fig:rgb}
\end{center}
\vspace{-0.6cm}
\end{figure}

Here, we present the discovery of a $\sim 250$ kpc X-ray tail behind the BIG based on the \cha\ and \xmm\ observations (Fig.~\ref{fig:rgb}). We assume a cosmology with $H_0$ = 70 km s$^{-1}$ Mpc$^{-1}$, $\Omega_m=0.3$, and $\Omega_{\Lambda}= 0.7$. At the A1367 redshift of $z=0.022$, $1^{\prime\prime}=0.445$ kpc.

\section{DATA ANALYSIS}
\label{s:obs}
Table~\ref{t:obs} lists the relevant \cha\ and \XMM\ observations.
We reduce the \cha\ observations with the \cha\ Interactive Analysis of Observation (CIAO; version 4.11) and Calibration Database (CALDB; version 4.8.2), following the procedures in \cite{2018MNRAS.481.4111G}. 
We process the \XMM\ data using the Extended Source Analysis Software (ESAS), as integrated into the \XMM\ Science Analysis System (SAS; version 17.0.0), following the procedures in \cite{2019MNRAS.484.1946G}.
We use the AtomDB (version 3.0.8) database of atomic data and the Solar abundance tables from \cite{2009ARA&A..47..481A}. The Galactic column density, $N_{\rm H}=1.91\times10^{20}\ {\rm cm}^{-2}$, is taken from the NHtot tool \citep{2013MNRAS.431..394W}.

\begin{table}
 \centering
  \caption{\cha\ and \XMM\ observations}
  \vspace{-0.2cm}
\tabcolsep=0.065cm  
  \begin{tabular}{@{}lccc@{}}
\hline\hline
Obs-ID & PI & Date & Exposure (ks)$^a$ \\ 
\hline
\cha \\
514 & Murray & 2000-02-26 & 40.5/38.4\\ 
4189 & Vikhlinin/Sun & 2003-01-24 & 47.5/42.4\\
17199 & Forman & 2015-01-30 & 37.9/37.7\\
17200 & Forman & 2015-11-05 & 39.6/39.4\\
17201 & Forman & 2016-01-31 & 61.3/61.1\\
17589 & Forman & 2015-01-31 & 24.8/24.8\\
17590 & Forman & 2015-02-01 & 36.6/36.4\\
17591 & Forman & 2015-02-03 & 34.9/34.9\\
17592 & Forman & 2015-02-08 & 25.7/25.7\\
18704 & Forman & 2015-11-08 & 27.0/26.4\\
18705 & Forman & 2015-11-14 & 23.8/23.6\\
18755 & Forman & 2016-02-03 & 48.4/48.4\\
\hline
\XMM \\
0005210101 & Fusco-Femiano & 2001-11-22 & 32.9(28.5)/26.1(14.9)\\
0061740101 & Forman & 2001-05-26 & 32.6(28.0)/29.1(17.4)\\   
0301900601 & Wolter & 2006-05-27 & 29.6(27.9)/19.6(13.0)\\
0602200101 & Finoguenov & 2009-05-27 & 24.6(20.7)/22.6(14.0)\\
0602200201 & Finoguenov & 2009-06-18 & 13.6(9.7)/12.5(7.0)\\ 
0602200301 & Finoguenov & 2009-11-25 & 22.1(19.5)/5.7(0.8)\\ 
0823200101 & Sun & 2018-06-01 & 71.6(69.8)/66.4(50.4)\\ 
0864410101 & Ge & 2020-12-05 & 40(38.1)/32.8(18.2)\\
\hline
\end{tabular}
\begin{tablenotes}
\item
$^a$: Total/clean exposure time, for \XMM, the exposure times are different from MOS and pn (in brackets).
\end{tablenotes}
\label{t:obs}
\end{table}

\begin{figure*}
\begin{center}
\centering
\includegraphics[width=0.45\textwidth,keepaspectratio=true,clip=true]{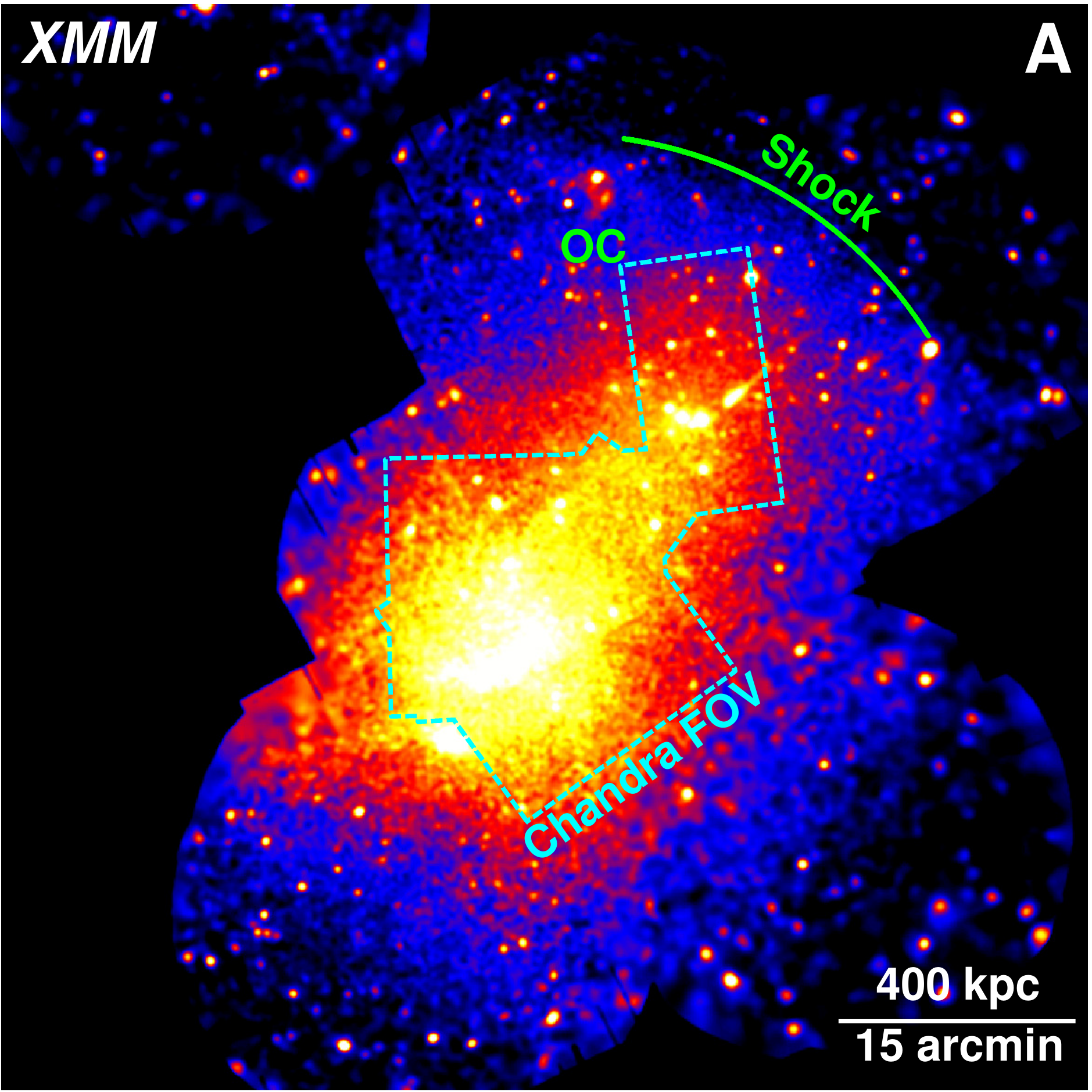}
\includegraphics[width=0.45\textwidth,keepaspectratio=true,clip=true]{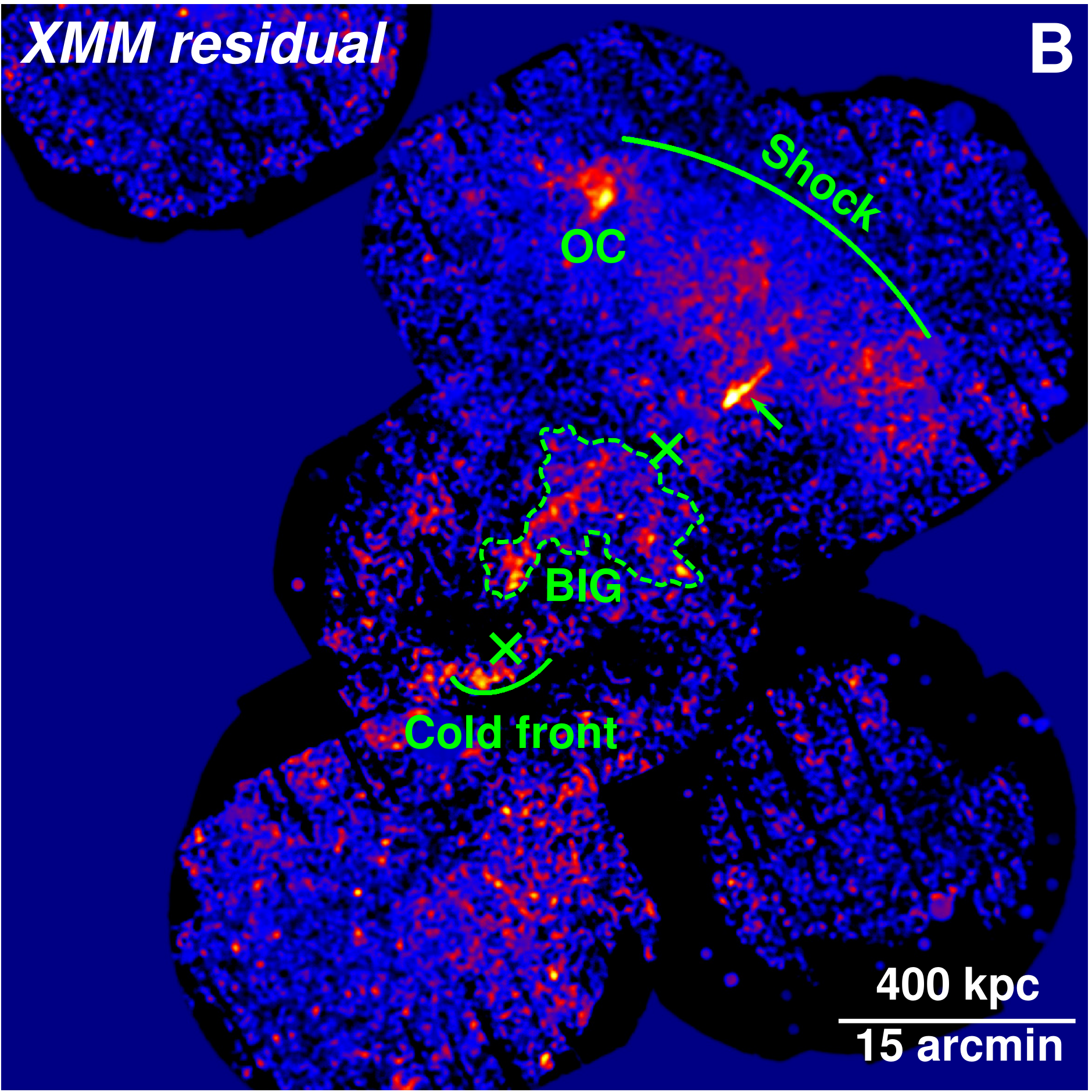}
\includegraphics[width=0.45\textwidth,keepaspectratio=true,clip=true]{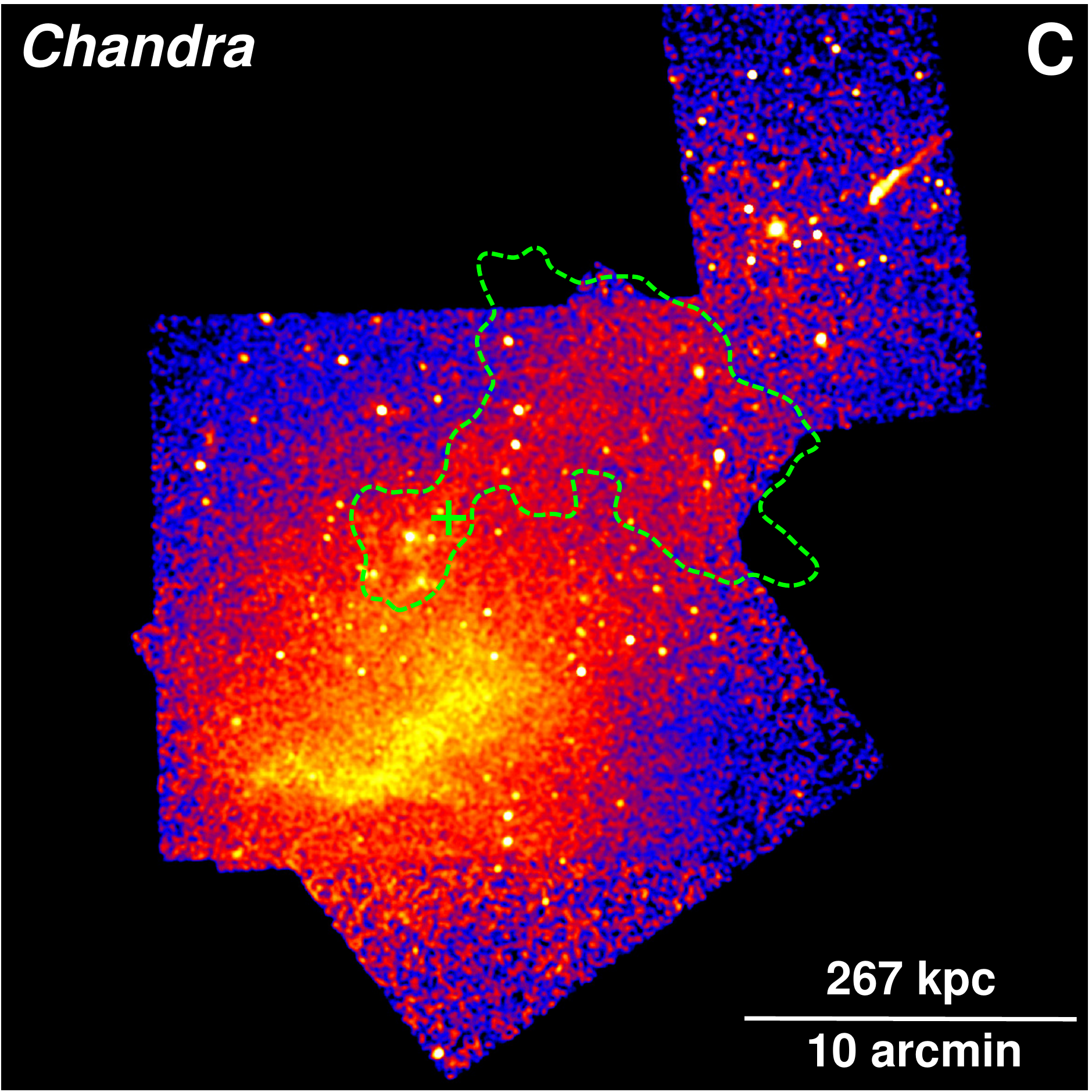}
\includegraphics[width=0.45\textwidth,keepaspectratio=true,clip=true]{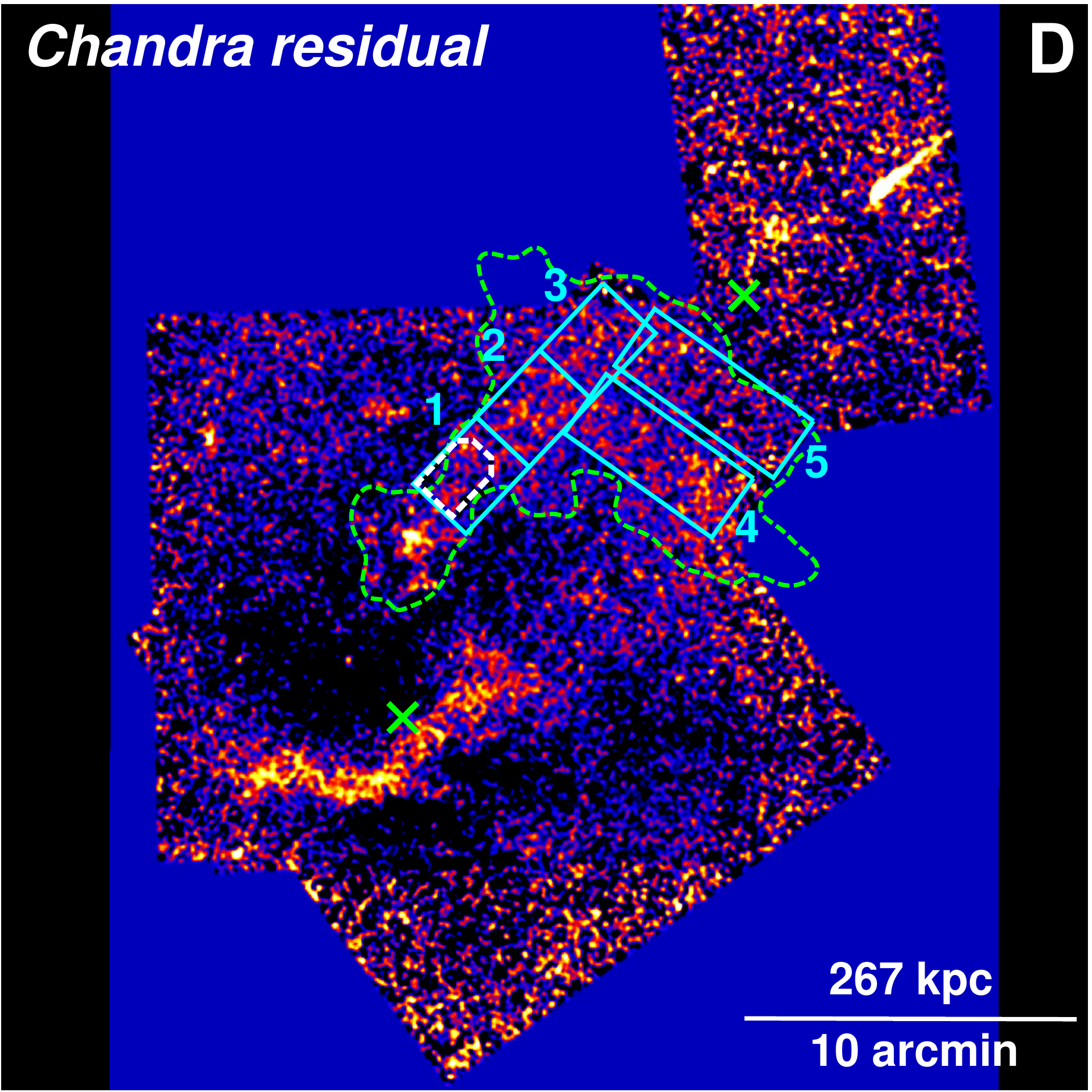}

\vspace{-0.2cm}
	\caption{X-ray images of A1367 and BIG.
	\textit{\textbf{(A)}}:\XMM\ combined 0.5-2.0 keV background subtracted, exposure corrected, and smoothed image. The cyan dashed polygon outlines the \cha\ FOV.
	Green arc marks a shock front \citep{2019MNRAS.486L..36G}.
	\textit{\textbf{(B)}}: \XMM\ 0.7-1.3 keV residual image with the best-fit ICM model and point sources subtracted. Notable residual features including: the cold front, the BIG (outlined by a dashed contour), UGC~6697 (pointed by a green arrow), and the orphan cloud (OC; \citealt{2021MNRAS.505.4702G}).
	Two green crosses mark the centres of two subclusters of A1367.
	\textit{\textbf{(C)}}: \cha\ combined 0.7-2.0 keV background subtracted, exposure corrected, and smoothed image. The dashed contour outlines the BIG. The green plus marks the K1 \citep{2002ApJ...578..842S} as the starting point of the BIG tail.
	\textit{\textbf{(D)}}: \cha\ residual image with the best-fit ICM model and point sources subtracted. The BIG tail is split into five regions marked as solid boxes, with spectral information in Table~\ref{t:fit}. The white dashed polygon is for the study of X-ray-\ha\ correlation. The somewhat different X-ray appearance of the front of the BIG between \cha{} and \XMM\ is caused by the different mask regions for their different PSF sizes.
	}
	 \label{fig:xray}
\end{center}
\vspace{-0.5cm}
\end{figure*}

\section{RESULTS}
\subsection{Imaging properties}
The BIG is projected between the centres of two subclusters of A1367 \citep{2021MNRAS.505.4702G}. The X-ray emission at its position is dominated by the cluster ICM emission.
While the X-ray excess behind the BIG is readily visible from both the \cha{} and \XMM{} images (Fig.~\ref{fig:xray}), the average cluster emission was subtracted to enhance the BIG feature.
We use two superimposed $\beta$-models to fit the ICM X-ray emission based on the \XMM\ data as detailed in \cite{2021MNRAS.505.4702G}. 
The X-ray tail of the BIG is revealed after subtracting the best-fit cluster ICM component from the original \XMM\ image as shown in Fig.~\ref{fig:xray}. 
After converting the count rates, in different X-ray bands, based on the ICM spectral model and response files from different instruments, the same best-fit ICM models are also subtracted from the \cha\ image to derive the residual \cha\ image in Fig.~\ref{fig:xray}.
\cite{2012MNRAS.421.1123C} found, for the Coma cluster, the large-scale fluctuations are model-dependent, while the perturbations smaller than several hundred kpc are model-insensitive.
This conclusion can be applied to A1367, as both clusters are at nearly the same distance.
To test the above argument, we also try the 2D elliptical $\beta$-models and patched $\beta$-models \citep{2015MNRAS.450.4184Z} to fit the X-ray emission of A1367, and conclude that the shown residual feature from the BIG is robust.

To increase the signal-to-noise ratio, we further merge the residual diffuse \cha\ and \XMM\ images. We firstly downgrade the resolution of the \cha\ image to match the \XMM\ image. We then rescale the \cha\ image to the \XMM\ image according to a ratio from their mean values within the overlapping field-of-view (FOV) marked as a dashed polygon in Fig.~\ref{fig:xray}(A). Finally, we add the rescaled \cha\ image to the \XMM\ image and take an average value within their overlapping FOV.
This method enhances the contrast between the real structure and background, because the background fluctuations are smoothed and suppressed. The resultant merged residual image is included in Fig.~\ref{fig:rgb}. The X-ray tail extends $\sim 250$ kpc in projection behind the extragalactic \hii\ knot K1 (\citealt{2002ApJ...578..842S,2006A&A...453..847C}) marked as a plus in Fig.~\ref{fig:xray}(C). 

\subsection{Spectral properties}
Background modeling is crucial to recover the relatively faint X-ray emission from the BIG tail. 
We extract on-source BIG tail spectra within the box regions in Fig.~\ref{fig:xray}(D) and off-source local background spectra from the \cha\ and \XMM\ data, respectively. We apply the double-background subtraction method to jointly fit the spectra from different instruments (e.g. \citealt{2019MNRAS.484.1946G}). Firstly, the instrumental non X-ray background (NXB) spectra are extracted from the stowed data sets for \cha, and from the filter-wheel closed data sets for \XMM. These rescaled NXB spectra are then loaded into {\tt XSPEC} as the background spectra. 
Secondly, the X-ray backgrounds, including the sky and ICM components, are properly modeled in {\tt XSPEC}.
The ICM component is 
fitted by an absorbed {\tt APEC} ($kT=3.9 {\rm\ keV}$, $Z=0.3\ Z\odot$).
Because the ICM emission varies across the on-source and off-source regions, their contributions are rescaled by the ratio from the same regions of a mock ICM image based on the above two $\beta$-model fitting.
The sky components are represented by an unabsorbed {\tt APEC} ($kT=0.12 {\rm\ keV}$, from the local hot bubble), 
an absorbed {\tt APEC} ($kT=0.21{\rm\ keV}$, Galactic background),
and an absorbed {\tt POWERLAW} (${\rm Index}=1.46$, AGN dominated background).
The best-fit parameters of the sky components are from a fit to the global spectra with a much larger coverage.
The sky contributions of on-source and off-source regions are linked by their sky solid angles.
The X-ray emission of the BIG tail in the on-source region is fitted with an additional absorbed {\tt APEC} model, whose normalization is fixed to zero in the off-source region.
The best-fit results are included in Table~\ref{t:fit}.

The X-ray signal is dominated by the ICM component. X-ray surface brightness fluctuations on scales of $\sim 200$ kpc are around 5\%, estimated from the ratio between the RMS of the residual image and the best-fit ICM model image. As shown in Table~\ref{t:fit}, we manually vary the ICM normalization by 5\%, and find that the BIG tail is still robustly detected.
We further split the BIG tail into five sub-regions in Fig.~\ref{fig:xray}(D). Their spectral fits are listed in Table~\ref{t:fit}. Their overall temperature is around 1 keV, and their metallicity is lower than the typical value in groups (0.2-0.4 Solar; \citealt{2021Univ....7..208G}).
The apparently low metallicity may suggest the intrinsically multi-$T$ gas in the stripped tail and/or non-equilibrium ionization (e.g. \citealt{2010ApJ...708..946S}).

\begin{table}
\caption{X-ray spectral information for the BIG.}
\vspace{-0.2cm}
\begin{center}
\begin{tabular}{lccc}
\hline
\hline
Region$^a$ & $kT$ ($Z$)$^b$ & C-stat/d.o.f \\
\hline
1-3 & $1.05\pm0.08\ (0.07\pm0.03)$ & 4834/4609 \\
1-3 (ICM 5\%$\downarrow$) & $1.17\pm0.11\  (0.07\pm0.03)$ & 4833/4609 \\
1-3 (ICM 5\%$\uparrow$) & $0.93\pm0.12\ (0.10\pm0.06)$ & 4850/4609 \\
\hline
1 & $0.96\pm0.10\ (0.15\pm0.09)$ & 3622/3389 \\
2 & $1.15\pm0.24\ (0.05\pm0.05)$ & 3573/3286 \\
3 & $1.05\pm0.16\ (0.07\pm0.04)$ & 3246/3086 \\
4 & $1.18\pm0.13\ (0.43\pm0.38)$ & 1133/1014 \\
5 & $1.13\pm0.19\ (0.19\pm0.11)$ & 1177/1058 \\
\hline
\end{tabular}
\end{center}
\vspace{-0.2cm}
\begin{tablenotes}
\item
$^a:$ The regions are shown in Fig.~\ref{fig:xray}(D). The main BIG tail is composed of regions 1-3. The BIG tail is robust even if ICM fluctuations (5\% level) are taken into account.
$^b:$ The best-fit $kT$ (keV) and metallicity (Solar) of the {\tt APEC} model.
\end{tablenotes}
\label{t:fit}
\end{table}

\section{DISCUSSION}
\subsection{Stripped multiphase gas of the BIG}
In Fig.~\ref{fig:rgb}, the cometary trails of the stripped hot gas behind the BIG resemble a cone. 
The BIG galaxies are at the cone tip and there is a certain inclination angle of the cone axis with respect to the plane of the sky.
The cone asymmetry may be due to projection effects if the BIG is infalling into A1367 with a non-zero impact parameter resulting in a curved path or tail and has a substantial velocity along the line of sight (e.g. \citealt{2017A&A...605A..25E}), which is very likely because the velocity of the BIG is $\sim 1700$ \km\ higher than A1367 (e.g. \citealt{2004A&A...425..429C}).
The innermost tail is relatively narrow and has \ha\ emission, which indicates that its gas is freshly stripped from the ISM of the BIG galaxies.
While the outermost tail is much wider and diffuse.
It may be stripped from the group hot intergalactic medium (hIGM) gas, which has lower thermal pressure and can be stripped first, it has been mixed with the A1367 ICM and becomes wider probably due to turbulent diffusion (e.g. \citealt{2017A&A...605A..25E}).
We can estimate the average density of hot gas from the {\tt APEC} normalization (e.g. \citealt{2021MNRAS.505.4702G}), assuming a uniform density distribution within a cone volume with an altitude of $\sim 250$ kpc and a circular base diameter of $\sim 250$ kpc. 
The resultant average electron gas density is $5.8\times 10^{-4}{\rm\ cm}^{-3}$,  the total hot gas mass is $6.8\times 10^{10}\ {\rm M}_\odot$, and the total X-ray bolometric luminosity is $3.8\times 10^{41}{\rm\ erg\ s}^{-1}$  in the cone.
If we focus on the bright main tail behind the BIG within regions 1-3 in Fig.~\ref{fig:xray}(D), we can also assume a uniform density distribution in a cylinder with a height of $\sim$ 220 kpc and a circular base diameter of $\sim 60$ kpc. The average electron gas density is $1.2\times 10^{-3}{\rm\ cm}^{-3}$, the total hot gas mass is $2.0\times 10^{10}\ {\rm M}_\odot$, and the total X-ray bolometric luminosity is $2.1\times 10^{41}{\rm\ erg\ s}^{-1}$ in the cylinder.

When stripped tails move in the ICM, thermal conduction and hydrodynamic instabilities (e.g. Rayleigh-Taylor and Kelvin-Helmholtz instabilities) can destroy the tails on a relatively short timescale.
However, observations suggest that both thermal conduction and hydrodynamic instabilities are highly suppressed, maybe due to magnetic fields (e.g. \citealt{2007PhR...443....1M}).
If we assume a tangential velocity in the plane of the sky of 800 \km\ for the BIG \citep{2017ApJ...839...65Y}, the stripped $\sim 250$ kpc long tail (in projection) has survived for at least 300 Myr. 
Therefore, the stripped tails can survive for a long time and produce gas inhomogeneities or clumping in ICM, which is important to cluster cosmology and baryon physics (e.g. \citealt{2015MNRAS.448.2971I,2021MNRAS.505.4702G}).

The warm gas tail behind the BIG, traced by \ha\ emission, extends 184 kpc $\times$ 45 kpc with a mass of $\sim 3\times 10^{9}\ {\rm M}_\odot$ for a volume filling factor of 1 \citep{2017ApJ...839...65Y}.
The cold gas tail traced by \hi\ has a mass of $\sim 7\times 10^{8}\ {\rm M}_\odot$ (BIG-NW region of \citealt{2006A&A...453..847C}).
Both the \ha\ \citep{2019MNRAS.484.2212F} and \hi\ \citep{2006A&A...453..847C} tails experience a progressive velocity decrease from the BIG. This suggests that the multiphase gas is decelerated relative to the BIG
due to the friction with the ICM after stripping.

We compare the X-ray and \ha\ surface brightness of the diffuse gas in the front part of the BIG tail as shown in Fig.~\ref{fig:xray}(D), where the robust H$\alpha$ flux from {\em MUSE} is available \citep{2019MNRAS.484.2212F}. The X-ray-to-\ha\ emission ratio is 4.0$\pm$1.4, which is consistent with the average ratio of $\sim$ 3.5 established from a sample of stripped tails in nearby galaxy clusters \citep{2021arXiv210309205S}. The correlation suggests the multiphase gas in tails originates from the mixing of the stripped ISM with the hot ICM. Further X-ray/H$\alpha$ comparison requires deeper H$\alpha$ data with good flux calibration throughout the BIG X-ray tail.
The role of magnetic fields in the stretched tail could also be important \citep[e.g.][]{2013MNRAS.436..526C}. 

\subsection{Nature of the BIG}
The generally conical RPS tail behind the BIG resembles other subcluster mergers in clusters like A2142 \citep{2017A&A...605A..25E}, A85 \citep{2015MNRAS.448.2971I}, and Coma cluster \citep{2021A&A...651A..41C}.
However, these infall subclusters are not star-bursting and have no report of \ha\ tails. By contrast, the BIG is a spiral-rich group with the highest density of star formation activity (total SFR $\sim 3\ {\rm M}_\odot{\rm\ yr}^{-1}$) in local clusters \citep{2006A&A...453..847C}.
Moreover, very few hIGM haloes have been observed in spiral-rich groups, and their origins are still under debate (e.g. \citealt{2014ApJ...793...74O}).
Interestingly, the NGC~4839 group is infalling into the Coma cluster along the direction of the Coma-A1367 filament,
while the BIG is infalling into the A1367 along a filament connecting  A1367 with the Virgo cluster (e.g. \citealt{2000ApJ...543L..27W}).

We can estimate the total mass of the BIG based on the virial theorem:
$M=\sigma^2 R/G\approx 1.0\times 10^{12} {\rm M}_\odot$, where $\sigma^2=3\sigma_r^2$ and $\sigma_r=170$ \km\ is the radial velocity dispersion of group members \citep{2006A&A...453..847C}; $R=50$ kpc is the size of BIG. 
Alternatively, we can also infer the mass from the X-ray properties. The $M-L$ relation (e.g. \citealt{2010ApJ...709...97L}) indicates $M \sim 1\times 10^{13} {\rm M}_\odot$, which may be underestimated as some gas has been stripped. The $M-T$ relation
(e.g. \citealt{2009ApJ...693.1142S}) indicates $M\sim 4\times 10^{13} {\rm M}_\odot$.
The mass from X-ray properties is a few $10^{13} {\rm M}_\odot$, which is one order of magnitude larger than from the virial theorem. 
However, $\sigma_r$ is probably underestimated because it is from only a few BIG members in a very compact region. 
If $\sigma$ increases by a factor of $\sim 3$ to $\sim 500$ \km\ and $M\sim \sigma^3$, 
the mass discrepancy from different methods can be alleviated. In fact, such a high $\sigma$ is suggested by the kinematics of ionized gas  \citep{2019MNRAS.484.2212F}, though the gas is disturbed by tidal interactions and RPS.
The mass of A1367 is $\sim 3\times 10^{14} {\rm M}_\odot$ from a mean $T \approx 3.5$ keV \citep{2009ApJ...693.1142S}. The merger mass ratio between A1367 and the BIG is $\sim 10$, which is a minor merger.

The perturbed stellar morphology/kinematics, tidal dwarfs, and a  complex of ionized gas filaments suggest that the BIG members are suffering from gravitational interactions (\citealt{2006A&A...453..847C,2019MNRAS.484.2212F}). The ISM stripped by tidal forces is further stripped by RPS when the BIG enters A1367. 
Multiphase gas including cold \citep{2006A&A...453..847C}, warm \citep{2019MNRAS.484.2212F}, and hot gas has been observed between BIG members (tidal origin) and behind the BIG (RPS origin).
Fig.~\ref{fig:rgb} shows a clear positional correlation between warm and hot gas, especially between the BIG members and in the main tail.
The winding feature of \ha\ tails in Fig.~\ref{fig:rgb} imprints the gas motion being tidally stripped out of galaxies \citep{2017ApJ...839...65Y}. 
Besides the BIG, CGCG~097-093 also has hybrid features of tidal interaction and RPS (e.g. \citealt{2017ApJ...839...65Y}). As shown in the lower-right corner of Fig.~\ref{fig:rgb}, its stripped X-ray and \ha\ \citep{2017ApJ...839...65Y} tails extend towards the NE with a length around 30 kpc.

\section{SUMMARY}
The BIG was discovered in previous \ha\ surveys of nearby clusters. Its galaxies are undergoing preprocessing via tidal interactions and RPS.
In particular, a long \ha\ tail of stripped gas is found behind the BIG. Based on the X-ray-\ha\ correlation in stripped tails \citep{2021arXiv210309205S}, we may expect a long X-ray tail behind the BIG.
However, the BIG is projected near the core of A1367, where the X-ray emission is dominated by the ICM.
Nevertheless, using \cha\ and \XMM\ data, after subtracting the ICM component of A1367, we discover a projected $\sim 250$ kpc X-ray tail behind the BIG. The stripped tail has survived for at least 300 Myr, and may contribute to ICM inhomogeneities or clumping.

The continual accretion of infalling groups is one of the main mechanisms to feed the growth of clusters.
Compared with relatively rare major cluster mergers, minor mergers between clusters and groups are more prevalent.
Combining future wide-field \ha\ and X-ray surveys,
more cases of infalling groups will be revealed to study the details of structure formation, ICM clumping, and galaxy preprocessing.

\footnotesize
\section*{ACKNOWLEDGEMENTS}
Support for this work was provided by the NASA grants 80NSSC19K0953 and 80NSSC19K1257 and the NSF grant 1714764.
M.F. has received funding from the European Research Council (ERC) under the European Union's Horizon 2020 research and innovation programme (grant agreement No 757535).
W.F. and C.J. acknowledge support from the Smithsonian Institution and the \cha\ High Resolution Camera Project through NASA contract NAS8-03060.
This research is based in part on data collected at \subaru\ Telescope, which is operated by the National Astronomical Observatory of Japan. We are honored and grateful for the opportunity of observing the Universe from Maunakea, which has the cultural, historical and natural significance in Hawaii.

\section*{DATA AVAILABILITY}
The \cha\ and \XMM\ raw data used in this paper are available to download at the HEASARC Data Archive website (https://heasarc.gsfc.nasa.gov/docs/archive.html).
The reduced data underlying this paper will be shared on reasonable requests to the corresponding authors.

\footnotesize
\bibliographystyle{mnras}
\input{ms.bbl}
\end{document}